# New Hybrid Maximum Power Point Tracking Methods for Fuel Cell using Artificial Intelligent

Mohammad Sarvi*, Masoud Safarishal**

Electrical Engineering Department, Imam Khomeini International University, Qazvin, Iran

*sarvi@eng.ikiu.ac.ir, **masoud.safarishaal66@gmail.com

**Abstract:** In this paper, two maximum power point tracking (MPPT) methods for Fuel Cell (FC) systems based on Adaptive Neuro-Fuzzy Inference Systems (ANFIS) and Imperialist Competitive Algorithm trained Neural Network (ICANN) are presented. The first operation voltage of the fuel cell corresponding to maximum power point is determined based on the data, and then the duty cycle of a DC/DC converter is adjusted using fuzzy logic controller to force the system that operates in conditions which match up with its maximum power point, in order to minimize the fuel consumption. The proposed systems and conventional fuzzy controller system are simulated in the MATLAB environment and results show acceptable operation under fast variation of conditions as well as normal conditions in minimum time.

# 1. Introduction

Modern power systems are becoming more complex and vulnerable to extreme events. These extreme events can seriously affect the operation of power systems, resulting in outages and even cascading failures [1]. As fossil fuel reserves are rapidly depleting, prices for fossil commodities are volatile, and fossil fuel has a negative environmental impact, most of the recent research focuses on the use of renewable forms of energy [2]. In the generation of energy, a future based on clean, environmentally friendly energy is important. Fuel cells are one of the cleanest and most efficient energy generation methods in the realm of renewable resources, according to several studies. Recent years have seen the development of many electrochemical devices (fuel cells) that use hydrogen and oxygen to generate electricity [3].

Today's Proton Exchange Membrane (PEM) fuel cell is one of the most commonly used types of fuel cells [4]. Because of low operating temperature, high power density, and fast startup, PEM fell cell is a promising candidate for residential and vehicular applications [5]. Fuel cells have nonlinear power-current (P-I) characteristics. Extensive discussions on the fuel cell P-I characteristic can be found in the literature [6]. On the other hand, their weakness is their low energy conversion efficiency. Therefore, to improve the performance of the FC, it must operate around the Maximum Power Point (MPP) which is influenced by temperature and membrane water content. The FC has an optimum operating point to generate the maximum power with certain membrane water content at a certain temperature. The P-I characteristic of the fuel cell is nonlinear, and the generating power changes with membrane water content and air temperature. Hence, the ability to extract the maximal power from a fuel cell is an important issue that must be considered for the optimal design of a fuel cell system. Some MPPT methods were proposed and investigated in [7-17]. In [6], these methods have been classified into three broad categories: offline methods, online methods, and hybrid methods. The problem addressed by the MPPT technique is to automatically find the maximum power point voltage or current at which a system should operate to obtain the maximum power output under given (and random) operating conditions. In this paper, in order to improve the speed of response to rapidly changing conditions, two new hybrid control methods are proposed for maximum power point tracking of the fuel cell system. In the first one, MPP tracking is done by the combination of Adaptive Neural Fuzzy Inference Systems (ANFIS) with the Fuzzy Logic (FL) controller. ANFIS is used to determine the optimal voltage of a fuel cell corresponding to maximum power. Temperature and membrane water content are the input of the ANFIS system and the reference voltage is its output. Afterward, the fuzzy logic controller is used to track MPP when tracking is done by adjusting the duty cycle of a DC/DC boost converter so that the fuel cell voltage remains at MPP operating point. The duty cycle of the power electronics converter is adjusted in order to push the operating point towards the MPP region as quickly as possible, thereby improving the transient response. Using fuzzy logic control has received significant attention over the last decade because it can deal with imprecise inputs, does not need an accurate mathematical model and can handle nonlinearity [19, 20]. ANFIS combines advantages of neural networks such as the learning power, with knowledge representation of a fuzzy inference system, so ANFIS can be applied to many complicated problems. Also, the advantages of ANFIS system over the traditional estimation methods are simple complementing of the model by new input parameters without modifying the existing model structure and automatic searching for the nonlinear connection between the inputs and outputs [21, 22]. In the second proposed method, Imperialist Competitive Algorithm (ICA) trained a neural network to determine the maximum power point. Machine learning become useful tools in a variety of applications such as model validation, modeling, and prediction [23, 24]. In recent years, methods based on NNs [25] and FL [26–28] have been successfully employed for the implementation of MPP searching. But finding the best weight factors in the network structure by trial and error takes a lot of time and sometimes may not be very accurate. Hence, to find the weighting factor, evolutionary algorithms have been proposed today [29-31]. In this article, ICA as a novel evolutionary algorithm used to train the neural network and achieve the best structure of the network [32]. ICA has been proposed by Atashpaz-Gargari and Lucas [33], in 2007, which has inspired from a socio-human phenomenon [34-36]. The ICA has been applied for optimizing the weighting factor of the neural network. The performance and accuracy of the proposed algorithm are investigated in different conditions and its results are compared with conventional fuzzy logic maximum power point tracking method. The main contribution of this paper is the presentation of two robust and reliable MPPT methods for tracking of MPP of FC under fast variation of operating conditions. The results of the proposed methods reduce oscillations and increased power yield under steady state conditions

The rest of this article is organized as follows: In section 2, the modeling of the PEM fuel cell system is presented. In section 3, the proposed MPP determination methods using ANFIS and

ICA trained the neural network, as well as conventional fuzzy methods, are presented. Fuzzy logic control algorithm which has been used for all three approaches is presented in section 4. In section 4, simulation results are presented, analyzed and discussed. Finally, the conclusions are presented in section 5.

## 2. Modeling of PEM fuel cell

In this article, a PEMFC is used to investigate the performance of the proposed maximum power point tracking methods. The mathematical model of the output voltage for PEMFC stack is formulated as follows [37-39]:

$$V_{cell} = E_{nernst} - V_{act} - V_{ohmic} - V_{con} \tag{1}$$

Where [40, 41]:

$$E_{nernst} = 1.229 - 8.5 \times 10^{-4}(T - 298.15) + 4.308 \times 10^{-5} T(\ln P_{H_2} + 0.5 \ln P_{O_2}) \tag{2}$$

$$V_{act} = -[\xi_1 + \xi_2.T + \xi_3.T.\ln(C_{O_2}) + \xi_4.T.\ln(i)] \tag{3}$$

$$C_{O_2} = \frac{P_{O_2}}{(5.08 \times 10^6) \times \exp(-498/T)} \tag{4}$$

$$V_{ohmic} = R_m.I \tag{5}$$

$$V_{con} = \frac{RT}{nF} \ln(1 - \frac{I}{i_L A}) \tag{6}$$

A PEMFC model has been developed using MATLAB/SIMULINK software. The fuel cell model can be shown as in Fig. 1. Also, Fig.2 shows the power current (P-I) characteristics of the fuel cell at different temperatures. As this Fig 2 shows, the MPP of FC is moved with changing of temperature and membrane water content.

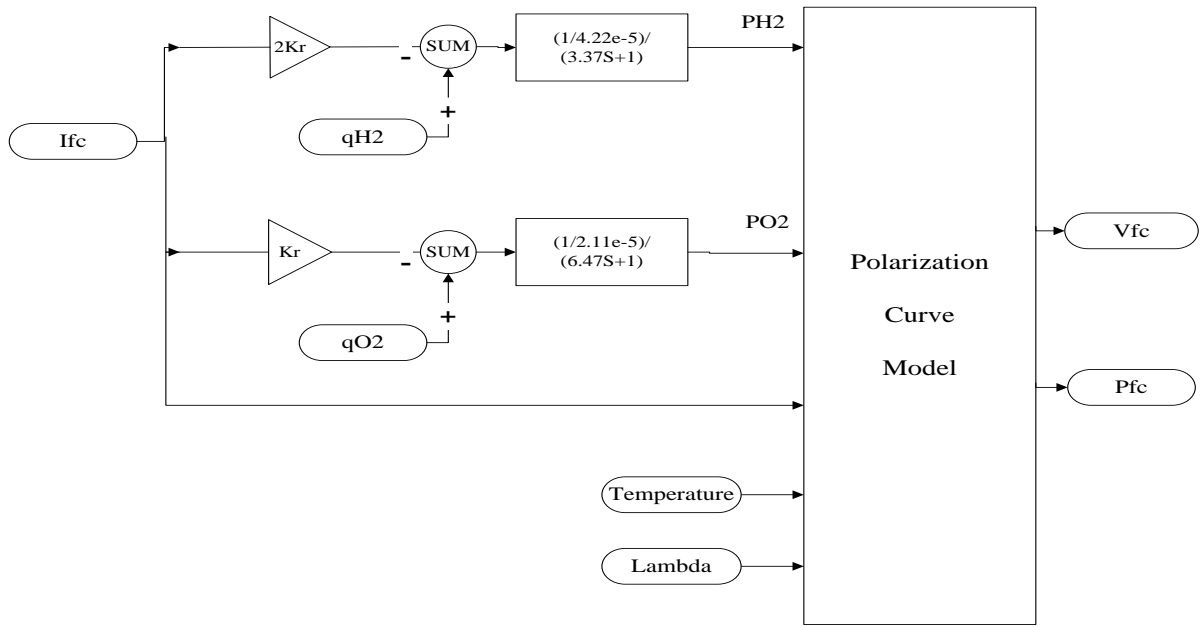

Fig. 1. PEMFC Model.

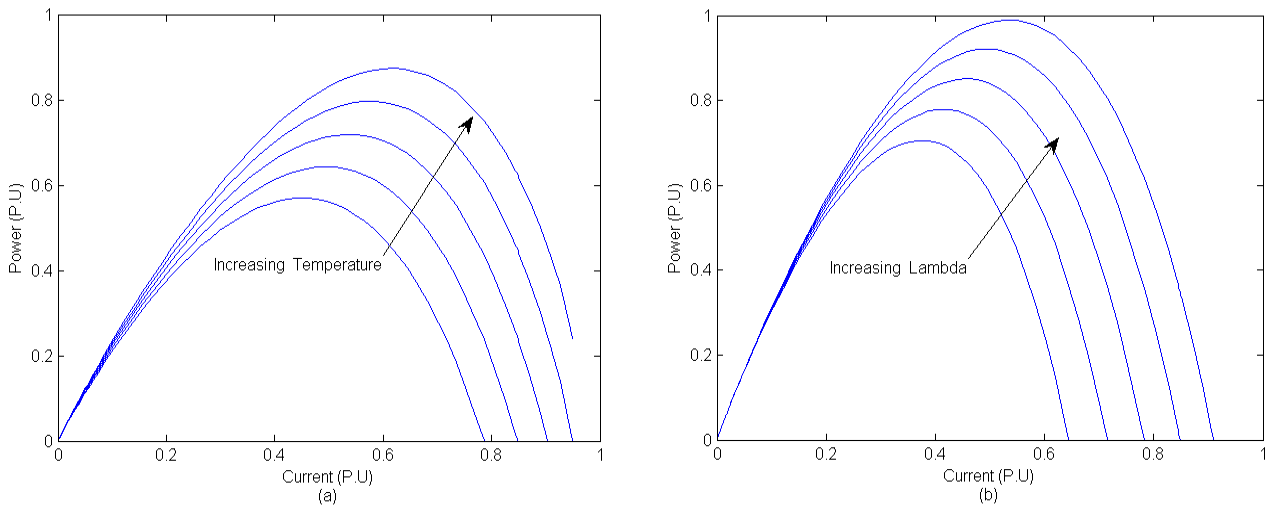

Fig.2. P-I characteristics of the fuel cell; a) at different temperature b) at different membrane water content.

## 3. The Proposed MPP Tracking Methods

In this section, two proposed MPP tracking methods as well as the conventional fuzzy MPP tracking method are presented. The first proposed method includes an ANFIS MPP determination and a fuzzy controller. In the second proposed MPP tracking approach, the Imperialist Competitive Algorithm trained Neural Network is used to determine the MPP of a fuel cell and the then fuzzy controller is used to track this point like the previous method. In order to compare the proposed methods with the conventional method, the third method (conventional fuzzy method) is presented. Authors proposed a conventional fuzzy MPP tracking method in [43]. Fig. 3 (a, b and c) shows the block diagram of three mentioned maximum power point tracking strategies.

## 3.1. The Proposed ANFIS based MPP Tracking

In this paper, ANFIS is used to find voltage corresponding to the maximum power point of FC (Vmax) in any operating conditions. An ANFIS is a neural network that is functionally equivalent to a fuzzy inference model which has the advantage of the learning capabilities of neural networks and modeling superiority of fuzzy systems simultaneously. In the proposed method, ANFIS determines fuel cell voltage corresponding to maximum power and the fuzzy controller adjusts fuel cell voltage to this determined voltage. The fuzzy controller is detailed in section 4. In order to investigate the performance of the proposed model, a system consisting of a PEM fuel cell, a DC/DC boost converter, a resistive load, and MPP tracker (including ANFIS and fuzzy controller) is considered as shown in Fig.3 (a). The PWM signal is used to turn on and off of the DC/DC boost converter switching element.

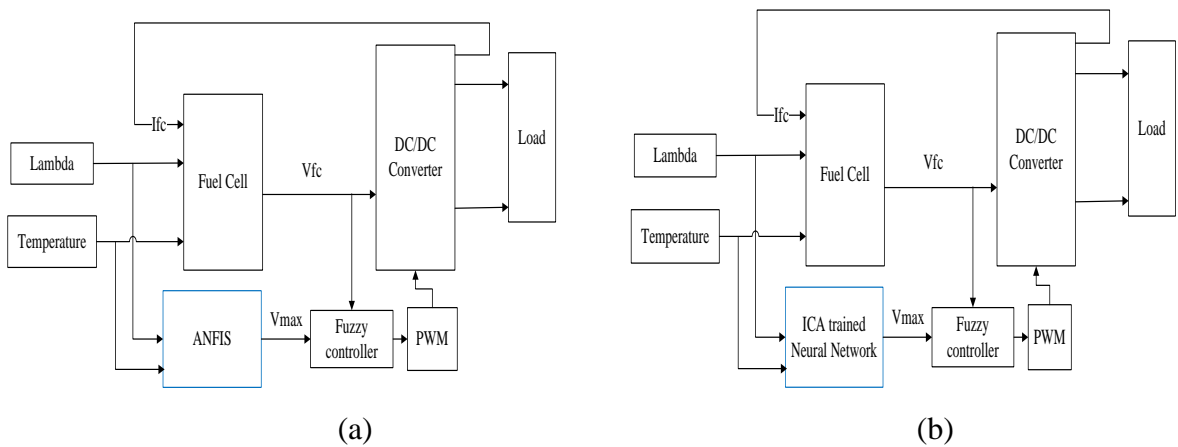

(a)                                          (b)

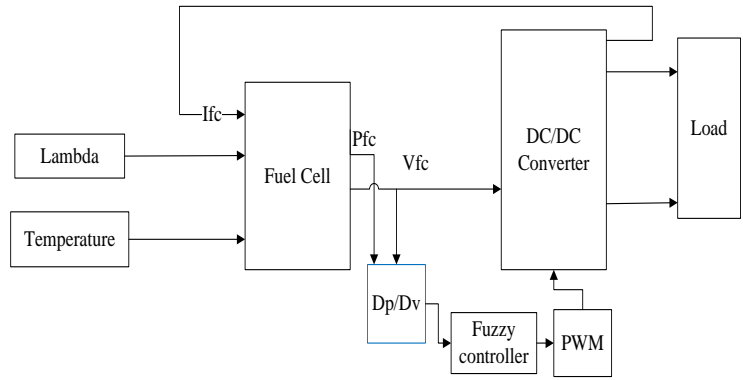

(c)

Fig.3. The block diagram of three MPP tracker for FC; a) the proposed ANFIS based MPP tracking, b) the proposed ICA based MPP tracking, c) the conventional fuzzy MPP tracking.

To train the ANFIS, a number of 250 data pairs are used. Temperature and cell membrane water content ($\lambda$) are the inputs of the ANFIS system, where Vmax with the above definition is ANFIS output. Three Gaussian membership functions are considered for each input. A separate data set, not included in the training set, is employed for verifying the ANFIS model generalization capabilities. The training and testing data are normalized. These normalized data are utilized as the inputs (operating conditions) and outputs (reference voltage) to train the ANFIS. After 70 epochs, training error reaches 0.001857 and correlation for testing data is equal to 0.99 percent. Results show ANFIS is a rapid and accurate prediction method.

*3.2. The Proposed ICA Trained Neural Network Based MPP Tracking*

Imperialist Competitive Algorithm (ICA) is applied to train MLP neural networks for enhancing the convergence rate and learning process. ICA has been proposed by Atashpaz - Gargari and Lucas [33]. ICA uses an evolutionary algorithm in order to optimize the weights of an MLP neural network. The ICA algorithm is a colonial competition inspired by the idea of human socio-political evolution. In the original algorithm, the number of colonial countries together with their colonial countries seeks to naturally find the general optimal point to efficiently solve the optimization problem. In this work, the numbers of initial countries are assumed to be Nc= 75 and the number of decades is Nd=65 and the number of initial imperialists is assumed to be Np= 8.

### 5.3.3. Conventional Fuzzy Method

The performance of the proposed methods is compared with the conventional fuzzy tracking method as well as actual data. As shown in Fig.4 at the MPP slope of power–voltage (P-V) curve is zero, at the left of the MPP slope is positive, and negative on the right. Fig.4 shows the power-voltage characteristic of a fuel cell. The inputs of MPP tracking controller are error (E), and error variation (CE), also output of MPPT controller is duty cycle variation of DC/DC converter (dD), where the error (E) and error variation are as following: $E(K) = \dfrac{\Delta P}{\Delta V} = \dfrac{P(k)-P(k-1)}{V(k)-V(K-1)}$

(7)

$$CE = E(K) - E(K-1) \tag{8}$$

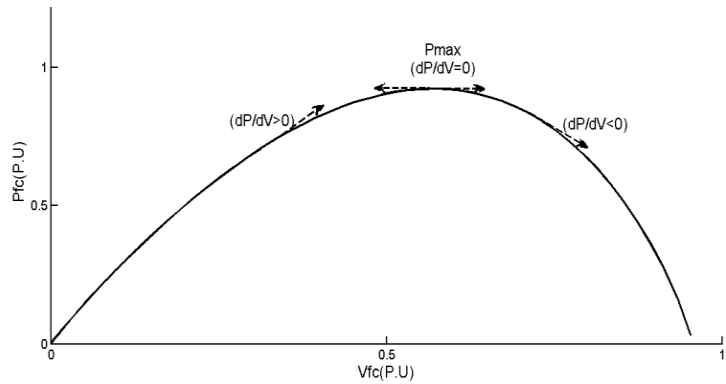

Fig.4. Power-voltage (P-V) characteristic of fuel cell.

For all three presented methods, the fuzzy logic controller has been used. The inputs of the fuzzy controller are an error (E), and error variation, (CE), and output of the fuzzy controller is duty cycle variation of DC/DC boost converter (dD) is. E is the difference between FC voltage and FC voltage corresponding to maximum power determined by ANFIS or ICA trained neural networks. Error and error variation for proposed methods are defined as the following:

$$E = V_{fc} - V_{max} \tag{9}$$

$$CE = E(K) - E(K-1) \tag{10}$$

Each of the membership functions of input and output variables has 7 triangular fuzzy subsets.

## 4. Simulations and Results

In order to investigate the performance and accuracy of the proposed MPP tracking methods, at the first step, a comparison between ANFIS and ICA trained neural network performances are presented. Thereafter simulations are performed for three different cases including of normal operating conditions and fast variation of the fuel cell temperature and the membrane water content, and then two proposed MPP tracking approaches are compared with conventional fuzzy MPP tracking method. Simulations are performed in the MATLAB/SIMULINK environment. A system consisting of a PEM fuel cell, a DC/DC boost converter, a resistive load, and MPP tracker is considered and simulated (as shown in Fig.3).

*5.1. Comparison between ANFIS and ICA trained neural network outputs*

In this section, two proposed estimator systems are compared with each other. Figs.5 (a) and (b) show ICA trained neural network output versus real output for train data and test data respectively. Figs.5 (c) and (d) show ANFIS output versus real output for train and test data. In these figures, the correlation of actual output with proposed methods outputs are displayed. Whatever the correlation rate is closer to 1, the estimator is more accurate. In addition, in order to compare the accuracy of the proposed methods, the Mean Squared Error (MSE) is computed. MSE of an estimator is one of many ways to quantify the difference between values implied by an estimator and the actual values of the quantity being estimated. The MSE of an estimator with respect to the estimated parameter (V) is given as follows: $MSE = \frac{1}{n}\sum_{i=1}^{n}(V_{Pi} - V_{Ti})^2$

(11)

Where VPi and VTi are ith elements of Vp and VT. Vp is a vector of n predictions values, and VT is a vector of the true values. The MSE and correlation value for both ANFIS and ICA trained neural networks are calculated for training and testing data. These results are shown in Table 1. As the results show, the accuracy of the model is acceptable in both methods, although ANFIS's results are slightly more accurate. Although, with more input data, ICA trained neural networks will perform better.

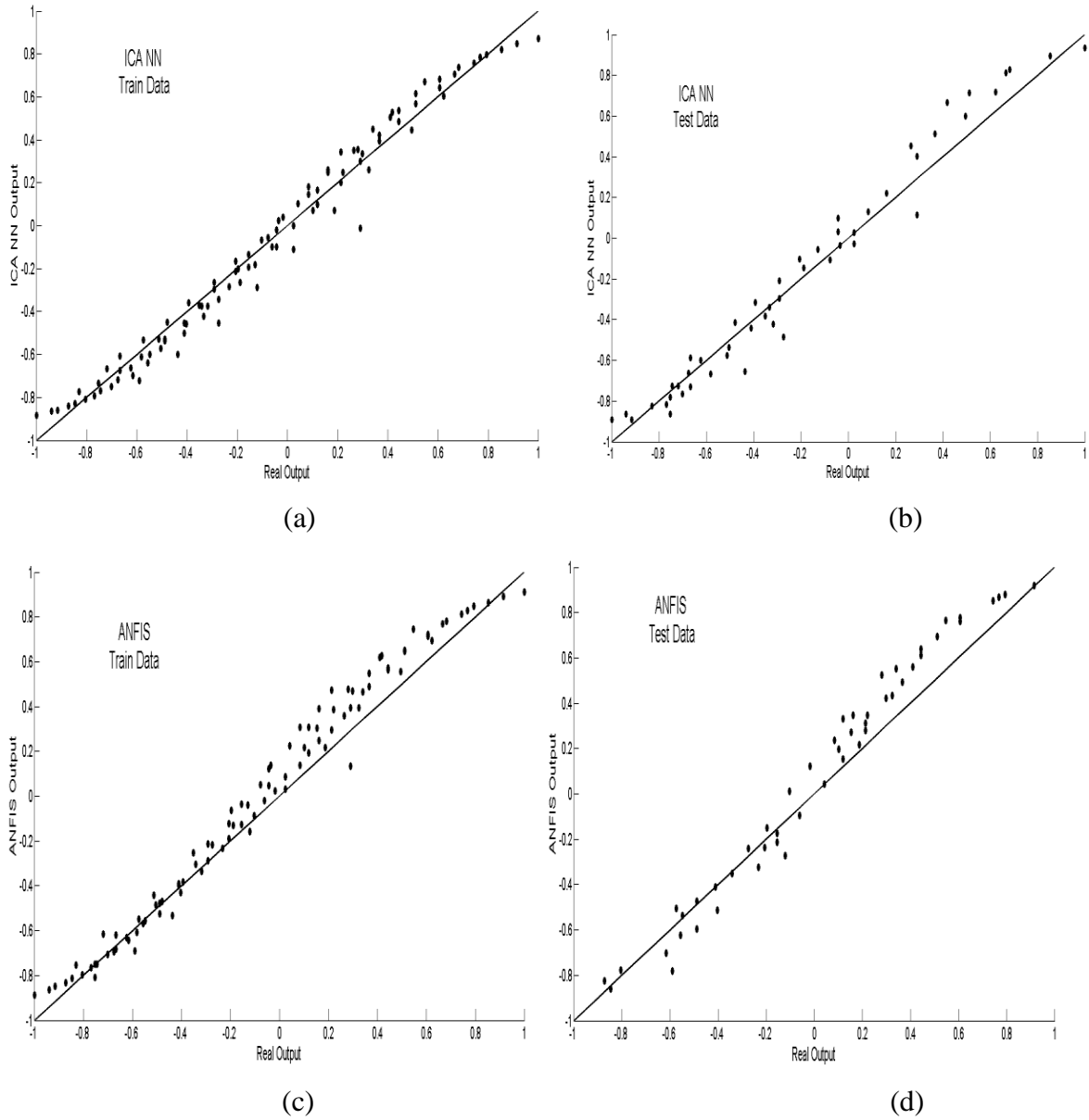

Fig.5. ICA trained neural network and ANFIS output versus real output for a) ICA NN train data b) ICA NN test data c) ANFIS train data and d) ANFIS test data.

Table 1. Comparison of ANFIS and ICA trained neural network.

|  | Training data |  | Testing data |  |
| --- | --- | --- | --- | --- |
|  | MSE | Correlation | MSE | Correlation |
| ANFIS | 0.0073 | 0.9906 | 0.0051 | 0.9882 |
| ICA trained NN | 0.0056 | 0.9894 | 0.0099 | 0.9862 |

*5.2. Normal Operating Conditions*

In this section, the values of membrane water content and temperature are fixed. In the first case, the value of membrane water content ($\lambda$) is assumed to be 12 and the value of temperature is assumed to be 40°C. In the second case, the values of membrane water content and temperature are assumed to be 13 and 55°C, respectively. Finally, in the third case of study, membrane water content and temperature are assumed to be 9 and 70°C, respectively. Table 3 shows the results for all three presented methods and real value at three different normal operation conditions. These results show that by using the proposed methods, the location of the maximum power point of the fuel cell is the nearest to the theoretical power as compared to the studied conventional method. Results also show ANFIS's results were slightly more accurate and its output is nearest to theoretical values as compared to ICA trained neural network. Table 2. Comparisons of the results of two proposed (ANFIS and ICA trained neural network) MPP tracking methods and conventional fuzzy MPP tracking, as well as actual value.

Table 2. Comparisons of the results ANFIS and ICA trained neural network normal operation

| Conditions<br>Methods | T=260°K, $\lambda$=12 | | T=280°K, $\lambda$=13 | | T=320°K, $\lambda$=9 | |
|---|---|---|---|---|---|---|
| | FC Maximum power(W) | Accuracy% | FC Maximum power(W) | Accuracy% | FC Maximum power(W) | Accuracy% |
| ANFIS | 3627 | 99.32 | 5305 | 98.91 | 6121 | 99.70 |
| ICA neural network | 3619 | 99.10 | 5321 | 99.20 | 6091 | 99.21 |
| Conventional Fuzzy | 3593 | 98.40 | 5203 | 97.01 | 6054 | 98.60 |
| Actual value | 3652 | 100 | 5364 | 100 | 6140 | 100 |

*5.3. Changing of the Fuel Cell Temperature*

In this case, a step change is applied to the temperature. The membrane water content is assumed constant and is equal to 12. The temperature changes from 50°C to 70°C at t=4 seconds, and then it changes from 70°C to 60°C at t=6 seconds, as shown in Fig.6 (a). The power corresponding to a maximum power point is tracked by two proposed and conventional fuzzy MPP tracking methods at different temperature and constant membrane water content, as shown in Fig.6 (b). These results show the proposed ANFIS and ICA neural network based MPP methods provide better performance (lower transient and more accurate response) than the conventional fuzzy MPP tracking method. On the other hand, the proposed methods determine the maximum power point faster and with more accuracy in comparison with conventional fuzzy MPP methods. Also, small settling time (Ts) and no overshoot are other good features of the proposed methods. Table 3 shows results for ANFIS and ICA neural network MPP tracking and conventional fuzzy MPP tracking, as well as an actual value.

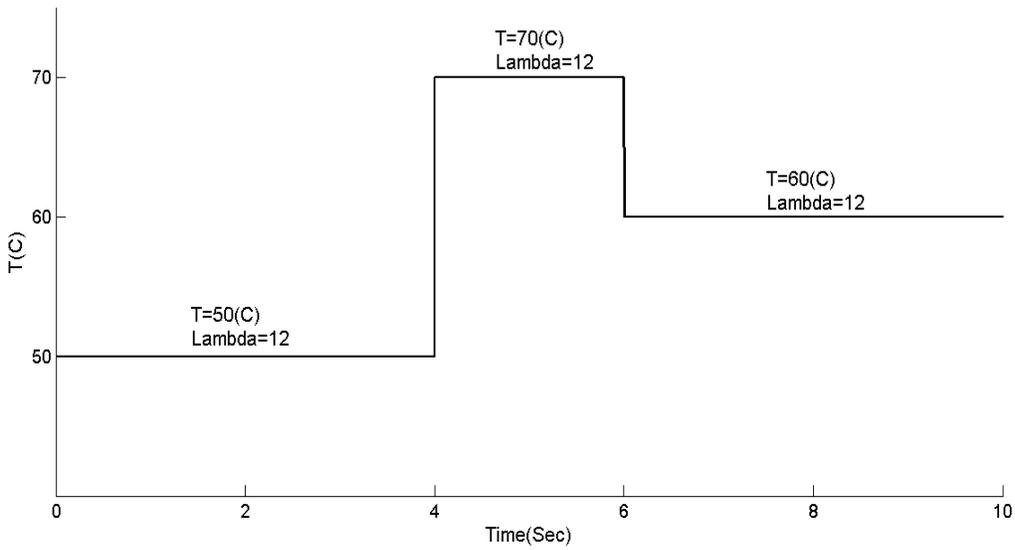

(a)

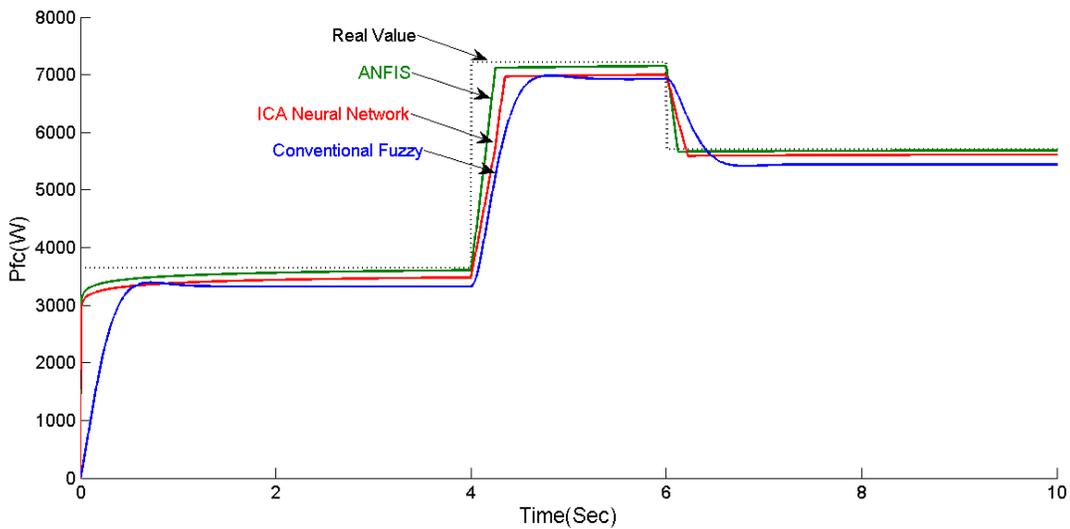

(b)

Fig.6. a) Temperature variations b) fuel cell power for proposed, conventional, and theoretical methods.

Table 3. Comparisons of proposed and conventional fuzzy methods at different temperature and constant cell membrane water content.

| Conditions<br>Methods | T=270°K, λ =12 | | | T=310°K, λ =12 | | | T=290°K, λ =12 | | |
|---|---|---|---|---|---|---|---|---|---|
| | $T_s$(s) | Accuracy% | FC Maximum power(W) | $T_s$(s) | Accuracy% | FC Maximum power(W) | $T_s$(s) | Accuracy% | FC Maximum power(W) |
| ANFIS | 0.12 | 98.63 | 3601 | 0.13 | 99.15 | 7144 | 0.06 | 99.68 | 5686 |
| ICA neural network | 0.15 | 98.00 | 3578 | 0.17 | 97.02 | 6991 | 0.11 | 98.36 | 5611 |
| Conventional Fuzzy | 0.26 | 93.83 | 3426 | 0.28 | 96.05 | 6921 | 0.27 | 95.38 | 5441 |
| Real Value | - | 100 | 3651 | - | 100 | 7205 | - | 100 | 5704 |

*5.4. Changing of the Fuel Cell Membrane Water Content*

In this section, the performance of the proposed MPP tracking methods under variation of cell membrane water content in constant temperature (55°C) is investigated. The membrane water content is changed from 9 to13 at t=4 seconds, and then it is changed from 13 to 11 at t=6 seconds, as shown in Fig.7 (a). The corresponding MPPs are changed as shown in Fig.7 (b). This figure shows the output power of the ANFIS and ICA neural network based MPP methods as well as fuzzy MPP trackers. Results show the proposed MPP methods have a better performance than conventional fuzzy MPP trackers when a rapidly changing of the MPP occurs. Fig.7 (b) highlights clearly the features of the proposed MPP tracking system, which improves the performance of the system by absorbing a higher power. The results in Table 4 show an enhancement of about 1% in energy absorbed by the proposed ANFIS an ICA neural network MPP tracker in comparison with a conventional fuzzy tracker.

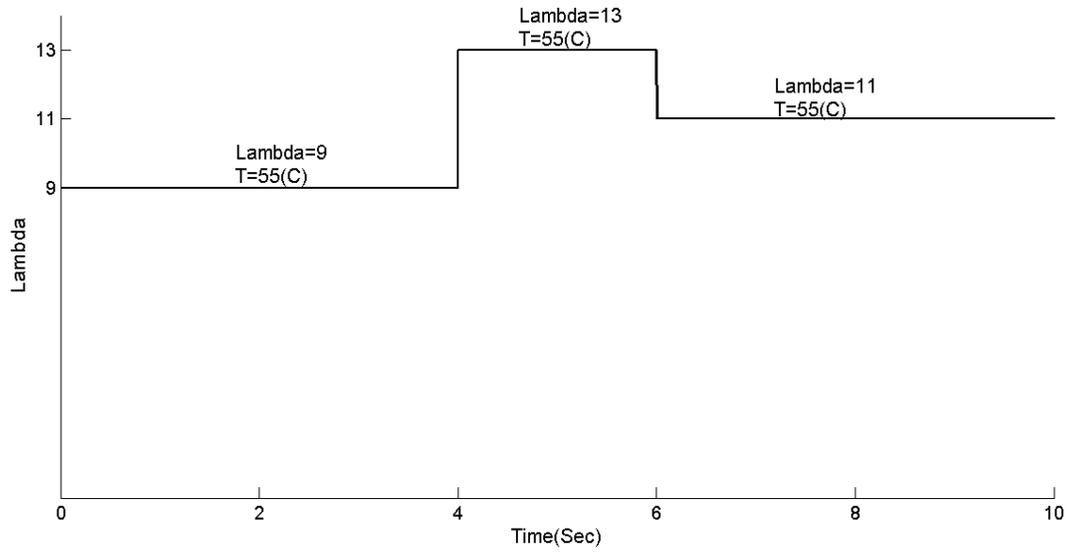

(a)

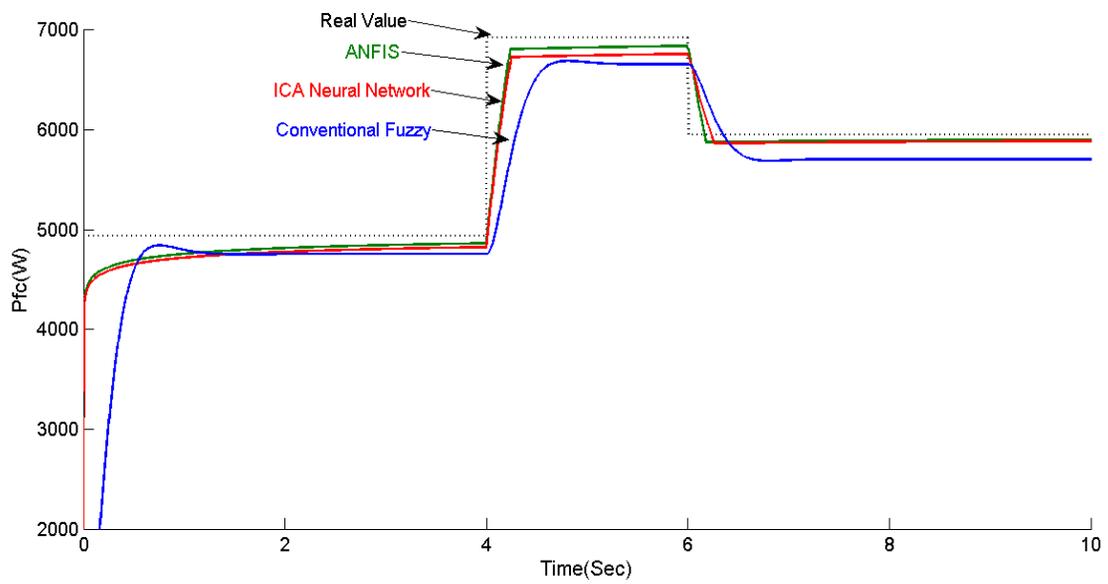

(b)

Fig.7. a) λ variation, b) fuel cell power for proposed and conventional fuzzy MPP tracking methods as well as real value.

Table 4. Comparison of proposed and conventional fuzzy MPP methods at different cell membrane water content (λ) and constant temperature.

| Conditions<br>Methods | T=300°K, λ=9 | | | T=300°K, λ=13 | | | T=300°K, λ=11 | | |
|---|---|---|---|---|---|---|---|---|---|
| | $T_s$(s) | Accuracy% | FC Max power(W) | $T_s$(s) | Accuracy% | FC Max power(W) | $T_s$(s) | Accuracy% | FC Maximum power(W) |
| ANFIS | 0.11 | 98.32 | 4858 | 0.12 | 98.71 | 6836 | 0.09 | 99.00 | 5896 |
| ICA neural network | 0.12 | 97.49 | 4817 | 0.12 | 97.55 | 6756 | 0.13 | 98.79 | 5883 |
| Conventional Fuzzy | 0.2 | 96.21 | 4754 | 0.25 | 96.12 | 6657 | 0.28 | 95.83 | 5707 |
| Real Value | - | 100 | 4941 | - | 100 | 6925 | - | 100 | 5955 |

## 6. Conclusion

In this paper, a new ANFIS and ICA trained neural network based MPP tracker was presented. The proposed methods combine a fuzzy MPP tracking with the artificial intelligence of ANFIS and neural network to speed up the procedure of reaching the accurate maximum power point of a fuel cell system under different conditions. ANFIS and ICA trained neural networks are used to determine FC voltage corresponding to maximum power (Vmax). A typical system consisting of a PEM fuel cell, a DC/DC boost converter, a resistive load, and MPP tracker (including ANFIS, ICA trained Neural Network or conventional fuzzy) is considered and simulated in MATLAB/SIMULINK. Simulations are performed for three different conditions including normal operating conditions and fast variation of the fuel cell temperature (when the water content is fixed) and the membrane water content (When the temperature is fixed). Results show, in comparison with the conventional fuzzy algorithm, proposed methods provide a better transient response because the proposed method determines the maximum power point faster and also gives smoother output power at steady state .

It is found that the results of the proposed MPP tracker are very close to the actual values over a wide range of temperature and membrane water content levels. Also, small settling time and no overshoot are the good features of the proposed MPP tracking methods. According to the comparisons on the actual results, it has been shown that the ANFIS system is more accurate than the other models.

**Nomenclature:**

| | |
|---|---|
| $E_{nernst}$ | Nernst Voltage (v) |
| $V_{act}$ | Activation Voltage (v) |
| $V_{Ohmic}$ | Ohmic Voltage (v) |
| $\xi_i$ (i =1,2,3,4) | Parametric Coefficients |
| $V_{Con}$ | Concentration Over voltage(v) |
| $P_{H2}$ | Hydrogen pressure (Pa) |
| $P_{O2}$ | Oxygen pressure (Pa) |
| $T$ | Temperature(K) |
| $C_{O2}$ | Concentration of Dissolved Oxygen($mol\,cm^{-3}$) |
| $R_m$ | Ohmic Resistance($\Omega$) |
| $r_m$ | Membrane Resistivity($\Omega cm$) |
| $A$ | Cell Active Area ($cm^2$) |
| $t_m$ | Membrane Thickness ($cm$) |
| $i_L$ | Limiting Current (A) |
| $F$ | Faraday constant, 96487 Charge (mol) |
| $\lambda_m$ | Membrane water content |